\documentclass[aip,amsmath,amssymb,preprint,floatfix]{revtex4-2}

\usepackage{graphicx}
\usepackage{bm}
\usepackage{booktabs}
\usepackage[colorlinks=true,linkcolor=blue,citecolor=blue,urlcolor=blue]{hyperref}

\newcommand{\kL}{k_{L}}
\newcommand{\kD}{k_{D}}
\newcommand{\LL}{L_{L}}
\newcommand{\vv}{\bm{v}}
\newcommand{\rr}{\bm{r}}
\newcommand{\zhat}{\hat{\bm{z}}}
\newcommand{\omhat}{\hat{\bm{\omega}}}

\begin{document}

\title{Exact turning invariants for projectile motion under quadratic drag and a
vertical-axis Magnus force}

\author{Pragyaan Gaur}
\email{pragyaangaur12@gmail.com}

\date{20 September 2026}

\begin{abstract}
A ball spinning about a vertical axis is pushed sideways by the Magnus force, so
its ground track curves. We show that when the lift and drag coefficients are
constant, the horizontal part of this problem is exactly solvable in elementary
functions. The step that makes this work is to write the horizontal velocity as a
complex number and to use path length rather than time as the independent variable.
Both aerodynamic forces are quadratic in speed, so the speed cancels and what
remains is a linear equation with constant coefficients in which gravity does not
appear. Three exact results follow from it. The heading turns at a fixed rate per
unit path length, so a half turn always costs the same distance whatever the launch
conditions and whatever the drag. The horizontal speed decays exponentially in the
turn angle at a rate fixed by the drag-to-lift ratio, which makes the horizontal
hodograph a logarithmic spiral. The radius of curvature of the ground track equals
the lift length times the cosine of the flight-path angle. The whole flight problem
also collapses to one scalar differential equation controlled by two dimensionless
numbers. We use this to settle a concrete question. Two identical balls thrown in
opposite horizontal directions with opposite spin can meet head-on when drag is
absent, on a one-parameter family we give in closed form, and they fail to meet once
drag acts. Every analytic claim is checked against machine-precision numerics, and
the code and data are openly available.
\end{abstract}


\maketitle

\section{Introduction}
\label{sec:intro}

The sideways deflection of a spinning ball is one of the most familiar everyday
demonstrations of fluid dynamics, and one of the oldest to be studied
quantitatively. Briggs measured the deflection of a baseball spinning about a
\emph{vertical} axis in a wind tunnel and found how it depends on spin rate and
speed \cite{Briggs1959}. Mehta's review is still the standard entry point to
sports-ball aerodynamics \cite{Mehta1985}, and measurements of the lift and drag
coefficients of spinning balls are now routine \cite{Nathan2008,KensrudSmith2018}.

The equations of motion for a spinning ball under gravity, quadratic drag and the
Magnus force are easy to write down, and they are almost always integrated
numerically. Bray and Kerwin, in the standard treatment of a soccer ball struck with
sidespin, put the position plainly: ``These equations have no closed form solutions
but can be solved numerically using a Runge--Kutta routine'' \cite{BrayKerwin2003}.
Nathan's reference study of baseball flight also uses fourth-order Runge--Kutta
\cite{Nathan2008}, as does the recent pedagogical treatment by Mencke \textit{et al.}
\cite{Mencke2020}. In the projectile-dynamics literature, Turkyilmazoglu and Altundag
obtain closed forms when either the quadratic drag or the Magnus term can be dropped,
and they describe the case in which both act together as highly nonlinear, with only
perturbative solutions available \cite{Turkyilmazoglu2020}. Chudinov's asymptotic
analysis of the same problem gives the velocity hodograph as an approximate implicit
formula \cite{Chudinov2024}. Reviews of projectile motion with drag
\cite{Lubarda2022,Chudinov2022} trace the analytic tradition back to Johann
Bernoulli's 1719 idea of parametrising planar motion by the tangent angle rather than
by time, which decouples the equations but leaves the solution implicit.

This paper points out that one physically natural special case of the system is
exactly integrable in elementary functions. When the spin axis is
vertical and $C_L$ and $C_D$ are constant, the horizontal motion separates completely
from gravity and can be solved in closed form. The invariants that come out are
simple, and as far as we can tell they have not been written down before.

Section~\ref{sec:model} fixes the model. Section~\ref{sec:reduction} derives the
central identity and reduces the full flight problem to a single scalar equation.
Section~\ref{sec:consequences} collects the invariants. Section~\ref{sec:rendezvous}
poses and answers a concrete geometric question that puts the machinery to work,
namely whether two counter-spinning balls thrown in opposite directions can be made
to collide head-on. Section~\ref{sec:numerics} describes the numerical verification,
Section~\ref{sec:discussion} discusses scope and an open problem, and
Section~\ref{sec:conclusion} concludes.

\section{Model}
\label{sec:model}

A ball of mass $m$ and radius $r$, with cross-sectional area $A=\pi r^2$, moves
through still air of density $\rho$ under gravity, quadratic drag and the Magnus
force,
\begin{equation}
\frac{d\vv}{dt} = -g\zhat
 - \frac{\rho C_D A}{2m}\,|\vv|\,\vv
 + \frac{\rho C_L A}{2m}\,|\vv|\,\bigl(\omhat\times\vv\bigr),
\label{eq:eom}
\end{equation}
where $\omhat$ is the unit spin vector. Throughout we take the spin axis vertical,
$\omhat=\pm\zhat$, which is the pure-sidespin configuration studied in
Refs.~\cite{Briggs1959} and \cite{BrayKerwin2003}. It is convenient to absorb the
coefficients into two inverse lengths,
\begin{equation}
\kL=\frac{\rho C_L A}{2m},\qquad \kD=\frac{\rho C_D A}{2m},\qquad
\mu\equiv\frac{\kD}{\kL}=\frac{C_D}{C_L},
\label{eq:kdef}
\end{equation}
and to give a name to the \emph{lift length},
\begin{equation}
\LL=\frac{1}{\kL}=\frac{2m}{\rho C_L A}.
\label{eq:LL}
\end{equation}
Equation~\eqref{eq:LL} is the exact analogue of the radius of a constant-$C_L$ turn
in aircraft performance, where the centripetal requirement and the available lift
both scale as $v^2$ and the speed cancels.

We write $\psi$ for the azimuth of the horizontal velocity, measured as an unwrapped
angle rather than through an inverse tangent, and $s$ for path length, so that
$ds=|\vv|\,dt$. The flight-path angle is $\gamma$, with $\tan\gamma=v_z/|w|$, where
$w$ is defined in Sec.~\ref{sec:reduction}. Sections~\ref{sec:reduction} and
\ref{sec:consequences} assume $C_L$ and $C_D$ constant, and
Sec.~\ref{sec:beyondconst} reports what happens when that assumption is relaxed.

\section{Exact reduction}
\label{sec:reduction}

\subsection{The central identity}

Combine the two horizontal components of the velocity into a single complex number,
\begin{equation}
w = v_x + i v_y .
\label{eq:wdef}
\end{equation}
Three observations make the horizontal problem linear. Gravity is vertical and
contributes nothing to $w$. Drag is antiparallel to $\vv$ and contributes
$-\kD|\vv|w$. With $\omhat=\zhat$ the Magnus term involves
$\omhat\times\vv=(-v_y,\,v_x,\,0)$, which is horizontal and which in complex notation
is exactly $iw$, a rotation by $90^\circ$. Putting these together,
\begin{equation}
\frac{dw}{dt} = \bigl(i\kL-\kD\bigr)\,|\vv|\,w .
\label{eq:wdot}
\end{equation}
The right-hand side is proportional to $|\vv|$, and this is what makes the change of
variable from time to path length pay off. Using $ds=|\vv|\,dt$, the speed cancels
identically and Eq.~\eqref{eq:wdot} becomes linear with constant coefficients,
\begin{equation}
\frac{dw}{ds} = \bigl(i\kL-\kD\bigr)\,w ,
\qquad\text{so that}\qquad
w(s) = w_0\,e^{-(\kD-i\kL)s} .
\label{eq:master}
\end{equation}
Equation~\eqref{eq:master} is exact. It holds with gravity acting throughout, at any
launch angle and for any ball, and it does not involve the vertical motion at all.
Everything that follows comes from it.

Two features are worth emphasising. First, the quadratic velocity dependence of
\emph{both} aerodynamic forces is essential, because it is the common factor $|\vv|$
that cancels under $ds=|\vv|\,dt$. A Magnus force linear in speed, of the kind that
arises for vortices in superfluids where the Magnus--Lorentz correspondence is
standard \cite{Sonin1997}, would not have this property, and the associated turning
radius would then depend on speed. Second, the reparametrisation is in the spirit of
Bernoulli's tangent-angle method \cite{Lubarda2022}, with the difference that here
the angle variable is an exact affine function of path length. That is what upgrades
the solution from implicit to explicit.

\subsection{Reduction to one scalar equation}
\label{sec:scalar}

Because $|w(s)|$ is now known in closed form, the whole flight problem reduces to a
single scalar equation. Take the turn angle $u\equiv\psi$ as the independent variable,
which by Eq.~\eqref{eq:master} means $u=\kL s$, and let
\begin{equation}
Q(u) \equiv \tan\gamma(u) = \frac{v_z}{|w|} .
\label{eq:Qdef}
\end{equation}
The Magnus force has no vertical component, so $\dot v_z=-g-\kD|\vv|v_z$. Dividing
this by $\dot\psi=\kL|\vv|$ gives $dv_z/du+\mu v_z=-g/(\kL|\vv|)$. Multiplying by the
integrating factor $e^{\mu u}$ and using $|w|=|w_0|e^{-\mu u}$ together with
$|w|=|\vv|\cos\gamma$ yields
\begin{equation}
\frac{dQ}{du} = -\lambda\,\frac{e^{2\mu u}}{\sqrt{1+Q^2}},\qquad
Q(0)=\tan\theta ,
\label{eq:reduced}
\end{equation}
with the single dimensionless group
\begin{equation}
\lambda = \frac{g}{\kL\,v_0^2\cos^2\theta} = \frac{g\LL}{v_0^2\cos^2\theta} ,
\label{eq:lambda}
\end{equation}
where $\theta$ is the launch elevation and $v_0$ the launch speed.
Equation~\eqref{eq:reduced} carries the whole problem. Given $Q(u)$, the heading is
$\psi=u$, the horizontal speed is $|w_0|e^{-\mu u}$, the total speed is
$|w_0|e^{-\mu u}/\cos\gamma$, and the position follows by quadrature. The
three-dimensional velocity equation \eqref{eq:eom} has been reduced exactly to one
scalar equation in the two parameters $\lambda$ and $\mu$.

Two functionals of the solution encode the geometry we shall need. Since
$dz/du=v_z/(\kL|\vv|)=\sin\gamma/\kL$, the ball returns to its release height after a
turn $\Psi$ if and only if
\begin{equation}
H \equiv \int_0^{\Psi}\sin\gamma(u)\,du = 0 .
\label{eq:closure}
\end{equation}
Similarly $d(x+iy)/du=w/(\kL|\vv|)=\cos\gamma\,e^{iu}/\kL$, so the launch-to-landing
chord is
\begin{equation}
\text{chord} = \frac{1}{\kL}\int_0^{\Psi}\cos\gamma(u)\,e^{iu}\,du .
\label{eq:chord}
\end{equation}
Equation~\eqref{eq:chord} is exact and is the key formula of
Sec.~\ref{sec:rendezvous}.

\section{Invariants}
\label{sec:consequences}

\subsection{Constant turning per unit path length}

Taking the argument of Eq.~\eqref{eq:master},
\begin{equation}
\frac{d\psi}{ds} = \kL = \text{const} ,
\label{eq:turnrate}
\end{equation}
so the heading advances by a fixed angle per unit distance travelled. Gravity, launch
angle, launch speed and $C_D$ are all absent from Eq.~\eqref{eq:turnrate}. Drag is
antiparallel to $\vv$ and so contributes nothing to $v_xa_y-v_ya_x$. It slows the ball
along the path without bending it. A heading change of $\Delta\psi$ therefore always
costs
\begin{equation}
S = \frac{\Delta\psi}{\kL} = \Delta\psi\,\LL ,
\label{eq:pathlength}
\end{equation}
and in particular a $180^\circ$ turn costs exactly $\pi\LL$ of path, whatever the
launch conditions and whatever the drag. For a baseball with $C_L=0.20$ this is
$\pi\LL=883.6$~m.

\subsection{Logarithmic-spiral hodograph}

Taking the modulus of Eq.~\eqref{eq:master} and eliminating $s$ using
Eq.~\eqref{eq:turnrate},
\begin{equation}
|w| = |w_0|\exp\!\bigl[-\mu\,\Delta\psi\bigr],\qquad \mu=\frac{C_D}{C_L} ,
\label{eq:spiral}
\end{equation}
so the horizontal velocity traces a logarithmic spiral whose pitch is set entirely by
the lift-to-drag ratio. At $\Delta\psi=\pi$ this becomes the compact statement
$|w_f|/|w_0|=e^{-\pi C_D/C_L}$. We stress that Eq.~\eqref{eq:spiral} governs the
\emph{horizontal} speed. The total speed obeys no such law, because gravity
re-accelerates the ball on descent and terminal velocity places a floor under $v_z$.
Using $e^{-\pi C_D/C_L}$ for $|\vv_f|/|\vv_0|$ underestimates the true ratio by more
than an order of magnitude at small $C_L/C_D$, as Table~\ref{tab:vf} shows.

Equation~\eqref{eq:spiral} inverts to
\begin{equation}
\frac{C_D}{C_L} = \frac{\ln\bigl(|w_0|/|w_f|\bigr)}{\Delta\psi} ,
\label{eq:measure}
\end{equation}
which suggests a measurement. The lift-to-drag ratio of a real ball can be read off a
single tracked trajectory by comparing the decay of its horizontal speed with the
swing of its heading. Gravity drops out of Eq.~\eqref{eq:measure} identically, so no
correction for the vertical motion is needed.

\subsection{Curvature of the ground track}

The radius of curvature of the horizontal projection is
$R_h=(ds_h/dt)/(d\psi/dt)=|w|/(\kL|\vv|)$, and $|w|/|\vv|=\cos\gamma$, so
\begin{equation}
R_h = \LL\cos\gamma .
\label{eq:curvature}
\end{equation}
The track is tightest where the ball is climbing or diving most steeply, and it is
flattest at the apex, where $\gamma=0$ and $R_h=\LL$ exactly. This gives a second
measurement, since the radius of curvature of the ground track at the apex is $\LL$
directly, without separate knowledge of $C_L$, $m$ or $A$. In the drag-free case
$\gamma$ runs from $+\theta$ to $-\theta$, so
\begin{equation}
\frac{R_{\max}}{R_{\min}} = \sec\theta ,
\label{eq:oval}
\end{equation}
which says exactly how far the track departs from a circle.

\subsection{Gravity is the sole source of non-circularity}

Setting $g=0$ in Eq.~\eqref{eq:reduced} gives $\gamma\equiv0$ and hence
$R_h\equiv\LL$, so the ground track is a perfect circle of radius $\LL$ whether or not
drag is present. This is a sharper statement than it may look, because drag can reduce
the speed by any factor without altering the geometry. Table~\ref{tab:circle} confirms
it numerically. A baseball that loses $94\%$ of its speed still traverses a circle
whose fitted radius agrees with $\LL$ to eleven significant figures.

\begin{table}[t]
\caption{Zero-gravity ground track, showing the fitted circle radius against $\LL$.
Drag changes the speed by more than an order of magnitude and leaves the radius
unchanged.}
\label{tab:circle}
\centering
\begin{tabular}{llllll}
\toprule
ball & drag & fitted radius (m) & $\LL$ (m) & rel.\ error & speed change (m/s) \\
\midrule
baseball & off & 281.267705304 & 281.267705306 & $5.5\times10^{-12}$ & $50\to50.000$ \\
baseball & on  & 281.267705317 & 281.267705306 & $3.8\times10^{-11}$ & $50\to3.200$ \\
frisbee  & off & 8.075877046   & 8.075877046   & $5.4\times10^{-12}$ & $50\to50.000$ \\
frisbee  & on  & 8.075877047   & 8.075877046   & $8.9\times10^{-11}$ & $50\to29.619$ \\
\bottomrule
\end{tabular}
\end{table}

\begin{figure}[t]
\includegraphics[width=\linewidth]{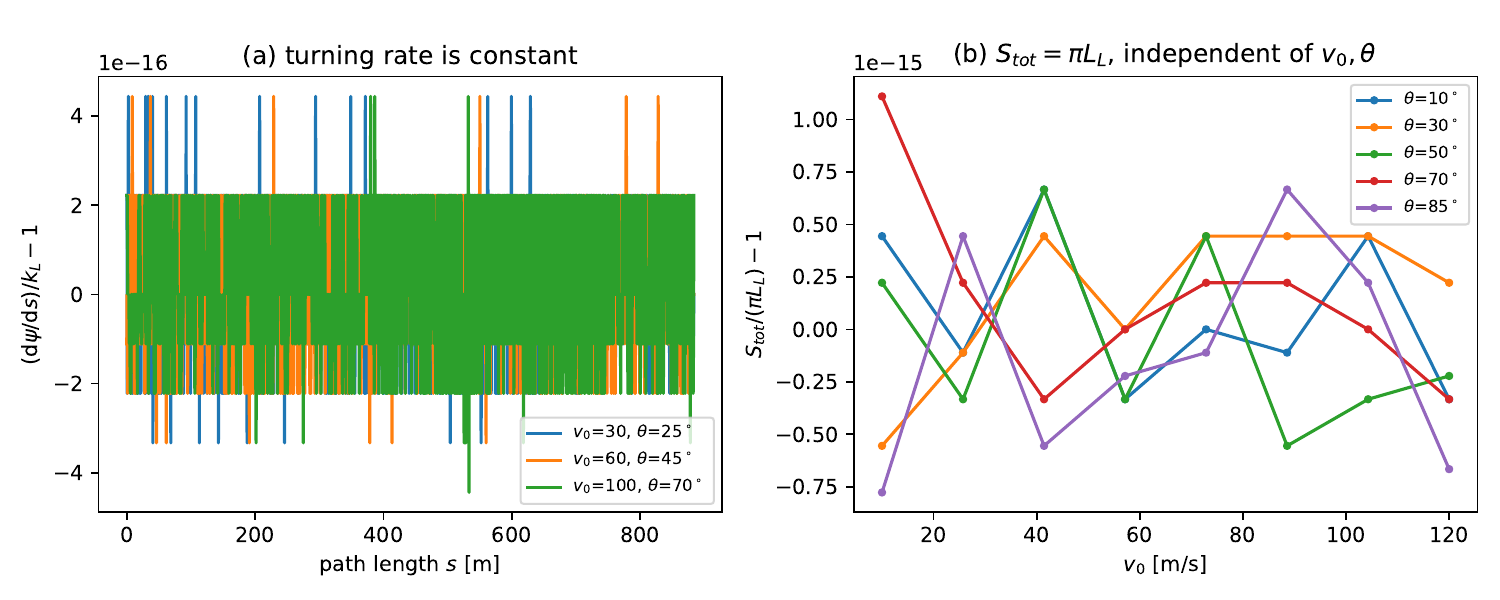}
\caption{Numerical verification of the turning invariants with drag off. Left: the
turning rate $d\psi/ds$ normalised by $\kL$, plotted along the path for three launch
conditions, with a deviation from unity that never exceeds $4.4\times10^{-16}$.
Right: the total path length required for a $180^\circ$ turn, normalised by $\pi\LL$,
swept over $v_0\in[10,120]$~m/s at five launch angles, with a deviation that never
exceeds $1.1\times10^{-15}$.}
\label{fig:invariants}
\end{figure}

\section{The counter-spinning rendezvous}
\label{sec:rendezvous}

\subsection{Statement of the problem}

We now put the machinery to work on a concrete question. A thrower at the origin
releases two identical balls at the same moment in opposite horizontal directions, one
spinning about $+\zhat$ and the other about $-\zhat$. Both curve in the same sense in
the horizontal plane. Can the launch conditions be chosen so that each ball turns
through exactly $180^\circ$ and the two collide head-on at the height of release?

Label the balls $A$ (launched along bearing $0$ with spin $+\zhat$) and $B$ (bearing
$\pi$ with spin $-\zhat$), and give them the same launch speed and the same elevation
angle. The configuration is invariant under reflection in the vertical plane at
bearing $90^\circ$ combined with the exchange $A\leftrightarrow B$, since a mirror
reflection reverses the horizontal velocity and negates the axial spin vector. Ball
$B$'s endpoint is therefore ball $A$'s endpoint reflected in $x$, and the two coincide
if and only if
\begin{equation}
\text{Re}\,(\text{chord}) = 0 ,
\label{eq:perp}
\end{equation}
that is, if and only if the launch-to-meeting chord is perpendicular to the launch
direction. By Eq.~\eqref{eq:chord} with $\Psi=\pi$, condition \eqref{eq:perp} reads
\begin{equation}
\mathcal{I} \equiv \int_0^{\pi}\cos\gamma(u)\,\cos u\,du = 0 .
\label{eq:Icond}
\end{equation}
The whole question is the sign of a single integral of the positive weight
$\cos\gamma$ against $\cos u$. The rendezvous succeeds precisely when that weight is
balanced about $u=\pi/2$.

\subsection{Without drag: an exact one-parameter family}

For $\mu=0$ Eq.~\eqref{eq:reduced} is separable, and integrating
$\sqrt{1+Q^2}\,dQ=-\lambda\,du$ gives
\begin{equation}
\tfrac12\bigl[Q\sqrt{1+Q^2}+\operatorname{arcsinh}Q\bigr]
 = \text{const} - \lambda u .
\label{eq:sep}
\end{equation}
The map $Q(u)\mapsto -Q(\pi-u)$ preserves the $\mu=0$ equation, so if the solution
satisfies $Q(\pi)=-\tan\theta$ then by uniqueness
\begin{equation}
Q(u) = -\,Q(\pi-u)
\label{eq:antisym}
\end{equation}
identically. Two things follow at once. First, $\sin\gamma$ is odd about $u=\pi/2$, so
the closure condition \eqref{eq:closure} holds automatically. Second, $\cos\gamma$ is
\emph{even} about $u=\pi/2$ while $\cos u$ is odd, so $\mathcal{I}=0$ exactly. Without
drag the chord is always exactly perpendicular to the launch direction, at every
launch angle. The rendezvous is therefore possible, and imposing $Q(\pi)=-\tan\theta$
fixes
\begin{equation}
\lambda = \frac{\tan\theta\sec\theta+\operatorname{arcsinh}(\tan\theta)}{\pi}
        = \frac{f(\theta)}{\pi\cos^2\theta} ,
\label{eq:lamfree}
\end{equation}
with $f(\theta)=\sin\theta+\cos^2\theta\,\ln(\tan\theta+\sec\theta)$. Combining
Eq.~\eqref{eq:lamfree} with Eq.~\eqref{eq:lambda} gives the launch speed in closed
form,
\begin{equation}
v_0 = \sqrt{\frac{\pi g \LL}{f(\theta)}} ,
\label{eq:v0}
\end{equation}
for \emph{any} $\theta$, which is a one-parameter family of solutions. Note that
$f(\theta)v_0^2/g$ is the classical arc length of a drag-free parabola, so
Eq.~\eqref{eq:v0} is just a restatement of Eq.~\eqref{eq:pathlength}. For a baseball,
Eq.~\eqref{eq:v0} gives $130.87$~m/s at $\theta=15^\circ$ and falls to a shallow
minimum of $84.99$~m/s near $\theta=56.5^\circ$ before rising again. At
$\theta=45^\circ$ the required speed is $86.89$~m/s and the meeting point lies
$515.4$~m away.

A detail worth recording, because it is easy to get wrong, is that at the meeting
point both balls are \emph{descending}. Their vertical velocities are common-mode and
cancel in the relative velocity. The closing speed is therefore
\begin{equation}
|\vv_A-\vv_B| = 2|w| = 2\cos\theta\,|\vv_f| ,
\label{eq:closing}
\end{equation}
rather than $2|\vv_f|$. We confirm this numerically to $6.5\times10^{-11}$ relative over
$\theta\in[10^\circ,80^\circ]$. At $\theta=60^\circ$ the closing speed equals the speed
of a single ball.

\subsection{With drag: the chord tips below perpendicular}

Once $\mu>0$ the antisymmetry \eqref{eq:antisym} is broken and $\mathcal{I}$ need not
vanish. Because the closure condition \eqref{eq:closure} determines $\lambda$ once
$\mu$ and $\theta$ are given, the chord bearing at closure is a function of $\mu$ and
$\theta$ alone. It cannot depend on $C_L$ and $C_D$ separately, nor on the mass, the
radius or the air density. This follows from the structure of Eq.~\eqref{eq:reduced}
rather than from inspection of the output, and it is confirmed numerically to six
decimal places for coefficient pairs that share a common $\mu$.

We evaluate the leading behaviour for small drag and small launch angle. Writing
$\beta$ for the chord bearing, a perturbative solution of Eq.~\eqref{eq:reduced},
given in the Appendix, yields
\begin{equation}
90^\circ-\beta \;=\; \Bigl(\frac{8}{3}-\frac{24}{\pi^{2}}\Bigr)\mu\,\theta^{2}
 \;+\; O(\mu^{2},\theta^{4})
\label{eq:deficit}
\end{equation}
in radians. The coefficient is $8/3-24/\pi^2=0.2349583$, and Richardson extrapolation
of the numerically computed deficit reproduces it to a relative accuracy of
$7\times10^{-6}$, as shown in the right panel of Fig.~\ref{fig:deficit}.

What matters here is the sign. The coefficient in Eq.~\eqref{eq:deficit} is positive
if and only if $8/3>24/\pi^2$, which is to say if and only if
\begin{equation}
\pi^2 > 9 .
\label{eq:pisq}
\end{equation}
Since $\pi^2=9.8696\ldots$, the deficit is positive, the chord tips below
perpendicular, and the rendezvous fails. The obstruction reduces to an elementary
inequality.

\subsection{A tempting argument that does not work}
\label{sec:tempting}

\begin{figure}[t]
\includegraphics[width=\linewidth]{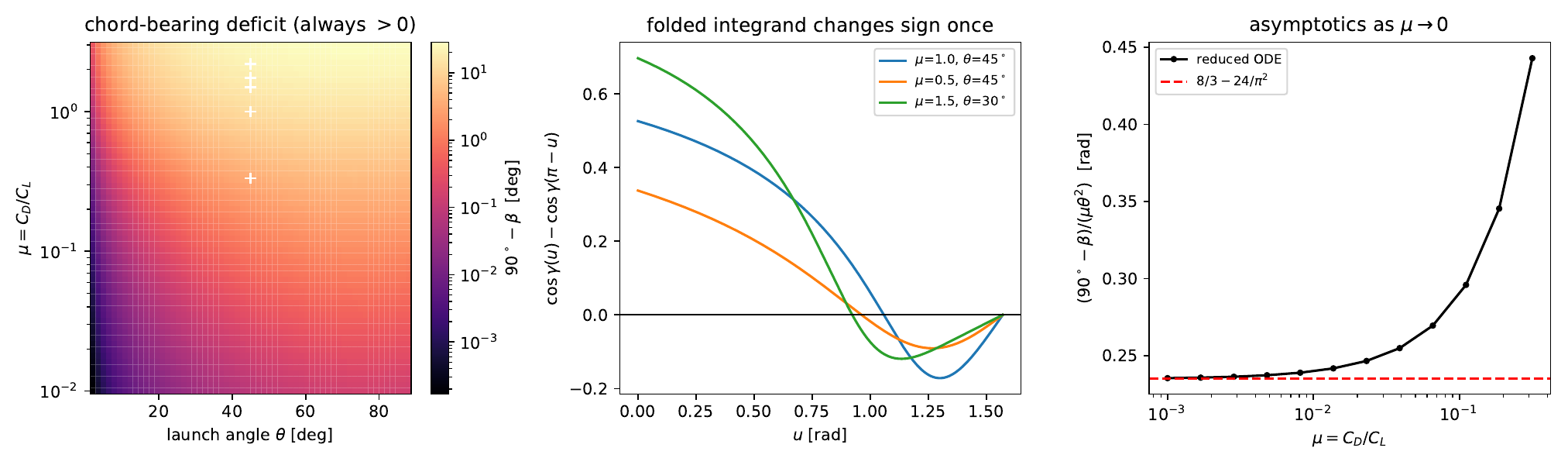}
\caption{The obstruction to the rendezvous, computed from the reduced equation
\eqref{eq:reduced}. Left: the chord-bearing deficit $90^\circ-\beta$ over $\mu$ and
$\theta$, strictly positive everywhere, with real sports balls marked at
$\theta=45^\circ$. Middle: the folded integrand of Eq.~\eqref{eq:folded}, which changes
sign once, showing that the positivity of $\mathcal{I}$ is not pointwise. Right:
convergence of the numerically computed deficit to the analytic coefficient
$8/3-24/\pi^2$ of Eq.~\eqref{eq:deficit} as $\mu\to0$.}
\label{fig:deficit}
\end{figure}

It is natural to try to prove $\mathcal{I}>0$ pointwise. Folding
Eq.~\eqref{eq:Icond} about $u=\pi/2$ gives
\begin{equation}
\mathcal{I} = \int_0^{\pi/2}\bigl[\cos\gamma(u)-\cos\gamma(\pi-u)\bigr]\cos u\,du ,
\label{eq:folded}
\end{equation}
and one might argue that with drag the descent is always steeper than the ascent at
matched turn angle, which would make the bracket positive everywhere. That argument
is false. Because drag makes the ascending arc longer than the descending arc, the
apex occurs at a turn angle \emph{greater} than $\pi/2$. We measure $u_a=0.543\pi$,
$0.572\pi$ and $0.635\pi$ for the three cases in Fig.~\ref{fig:deficit}. For $u$
slightly below $\pi/2$ both $u$ and $\pi-u$ then lie on the ascent, where $|\gamma|$ is
decreasing, so the ordering reverses. The bracket in Eq.~\eqref{eq:folded} changes sign
exactly once, as the middle panel of Fig.~\ref{fig:deficit} shows. The positivity of
$\mathcal{I}$ is genuinely an integral statement, in which a large positive
contribution near launch outweighs a smaller negative one near the midpoint. We record
this explicitly because the pointwise argument looks convincing at first.

\subsection{Numerical survey and consequences}

Beyond the asymptotic regime we evaluate the deficit directly from
Eq.~\eqref{eq:reduced} over a $60\times60$ grid spanning $10^{-2}\le\mu\le3$ and
$2^\circ\le\theta\le88^\circ$, which brackets every ball in Table~\ref{tab:balls}. The
deficit is strictly positive throughout, with a minimum of $1.67\times10^{-4}$ degrees
at the corner $\mu=10^{-2}$, $\theta=2^\circ$, where both small parameters vanish
together. This is the left panel of Fig.~\ref{fig:deficit}. We therefore state the
following.

\medskip
\noindent\textit{Result.} Take constant $C_L$ and $C_D$ with $C_D>0$ and a vertical
spin axis. Two balls released in exactly opposite horizontal directions with opposite
spin, at equal launch speeds and equal elevation angles, do not collide, at any launch
speed or launch angle. This is proved for small $\mu\theta^2$ by
Eq.~\eqref{eq:deficit} and established numerically over the physically relevant
domain. A proof valid for all $\mu$ and $\theta$ remains open, as does the
corresponding question for unequal launch conditions, which we discuss in
Sec.~\ref{sec:open}.
\medskip

The practical size of the failure is not small. The two balls miss by
$2|\text{chord}|\cos\beta$, which for the closure solutions of Table~\ref{tab:balls}
is $1.0$--$1.9$~m for a frisbee, $7.6$--$9.4$~m for a table tennis ball and $30$--$46$~m
for a soccer ball.

We note in passing that this changes the count of conditions. The two residuals
``return to launch height'' and ``$\psi=\pi$'' are two ways of writing the single
scalar statement that two events coincide, over the three unknowns $v_0$, $\theta$ and
$t^*$. The corresponding $2\times2$ Jacobian is numerically rank one, with a
singular-value ratio of $4.0\times10^{-9}$. The physical problem still becomes
genuinely two-dimensional once drag is present, through the independent condition
\eqref{eq:perp}, which without drag is satisfied automatically.

\section{Numerics and validation}
\label{sec:numerics}

The full system is integrated as an eleven-component state
$[\rr,\vv,\bm{\omega},\psi,s]$, carrying $\psi$ and $s$ as state variables so that the
heading unwraps correctly past $\pm\pi$ without any inverse-tangent differencing.
Integration uses explicit Runge--Kutta 4(5) with relative tolerance $10^{-10}$ and
event detection for the two closure events, never fixed steps
\cite{Virtanen2020,Harris2020}. Parameter sweeps use a batched fixed-step integrator.
Figures are produced with Matplotlib \cite{Hunter2007}.

Table~\ref{tab:verify} summarises the verification. The reduced equation
\eqref{eq:reduced} reproduces $\gamma(\psi)$ from the full three-dimensional
integration to better than $10^{-8}$~rad across four balls and three launch angles,
and the reduced chord bearing agrees with the full simulation to better than $10^{-4}$
degrees. The invariants \eqref{eq:turnrate} and \eqref{eq:pathlength} hold to machine
precision. Equation~\eqref{eq:turnrate} holds equally well with drag active, which
confirms that drag cannot torque the azimuth.

\begin{table}[t]
\caption{Verification of the analytic results against direct integration. Deviations
are relative unless stated otherwise.}
\label{tab:verify}
\centering
\begin{tabular}{lll}
\toprule
quantity & condition & max deviation \\
\midrule
$d\psi/ds=\kL$                        & drag off & $4.4\times10^{-16}$ \\
$d\psi/ds=\kL$                        & drag on  & $6.7\times10^{-16}$ \\
$S_{\rm tot}=\pi\LL$                  & drag off, $v_0\times\theta$ sweep & $1.1\times10^{-15}$ \\
$|w|=|w_0|e^{-\mu\Delta\psi}$         & drag on, four balls & $<10^{-9}$ \\
$R_h=\LL\cos\gamma$                   & drag on and off & $<10^{-12}$ \\
$g=0$ track is a circle of radius $\LL$ & drag on and off & $<10^{-9}$ \\
reduced ODE vs.\ full 3-D $\gamma(\psi)$ & drag on, four balls & $<10^{-8}$~rad \\
$\lambda$ from Eq.~\eqref{eq:lamfree} & drag off & $3.5\times10^{-12}$ \\
$v_0$ from Eq.~\eqref{eq:v0}          & drag off & $2\times10^{-14}$ \\
closing speed $=2\cos\theta|\vv_f|$   & drag off & $6.5\times10^{-11}$ \\
$R_{\max}/R_{\min}=\sec\theta$        & drag off & $<10^{-6}$ \\
\bottomrule
\end{tabular}
\end{table}

\subsection{Beyond constant coefficients}
\label{sec:beyondconst}

The results of Secs.~\ref{sec:reduction} and \ref{sec:consequences} require $C_L$ and
$C_D$ constant. Two remarks bound their scope. First, under constant $C_L$ the Magnus
term depends on the spin \emph{direction} only, so the spin magnitude is exactly inert.
Trajectories for $\omega=50$, $200$ and $800$~rad/s coincide to twelve digits. Any
question about sensitivity to spin rate is therefore empty until $C_L$ is allowed to
depend on the spin parameter $S=r\omega/|\vv|$. With the saturating form
$C_L=1/(2+1/S)$, a $10\%$ spin mismatch between the two balls produces a $1084$~m miss
at the drag-free $45^\circ$ solution, which is larger than the chord itself. Second, we
have implemented a smoothed drag-crisis model in which $C_D$ falls from $0.47$ to
$0.15$ across $\mathrm{Re}\sim2$--$4\times10^5$. With $C_D$ varying, $|w|$ no longer
follows Eq.~\eqref{eq:spiral} exactly, although Eq.~\eqref{eq:turnrate} survives so
long as $C_L$ is constant, since drag never contributes to the turning rate whatever
its magnitude.

\subsection{Real balls}

Table~\ref{tab:balls} collects the closure solutions with drag active, using
representative coefficients \cite{Mehta1985,Nathan2008,KensrudSmith2018}. Only three of
the six objects can complete a $180^\circ$ turn at all below $150$~m/s, because the
required path $\pi\LL$ must be covered before drag exhausts the ball. A baseball
manages $0.504\pi$ of turn at best, falling short by $0.496\pi$, and a solution exists
mathematically only near $v_0=616$~m/s, which is Mach~$1.8$ and outside the validity of
an incompressible quadratic-drag model. In view of Sec.~\ref{sec:rendezvous}, we
emphasise that the entries in Table~\ref{tab:balls} are single-ball $180^\circ$
closures and are \emph{not} rendezvous solutions.

\begin{table}[t]
\caption{Closure of a single $180^\circ$ turn with drag active and $v_0$ capped at
$150$~m/s. The minimum-$v_0$ member of the $\theta$-family is shown. Coefficients are
representative constants rather than measurements.}
\label{tab:balls}
\centering
\begin{tabular}{lcccccc}
\toprule
object & $C_L/C_D$ & closes & $\theta$ & $v_0$ (m/s) & $t$ (s) & $v_f/v_0$ \\
\midrule
frisbee   & 3.00  & yes & $64^\circ$ & 19.85  & 2.813 & 0.590 \\
soccer    & 1.00  & yes & $76^\circ$ & 124.20 & 9.583 & 0.211 \\
table tennis & 0.667 & yes & $80^\circ$ & 90.95  & 3.966 & 0.096 \\
golf      & 1.00  & no  & --- & --- & --- & $\psi_{\max}=0.804\pi$ \\
baseball  & 0.571 & no  & --- & --- & --- & $\psi_{\max}=0.504\pi$ \\
tennis    & 0.455 & no  & --- & --- & --- & $\psi_{\max}=0.565\pi$ \\
\bottomrule
\end{tabular}
\end{table}

\begin{table}[t]
\caption{Residual speed at closure, comparing the exact horizontal law
\eqref{eq:spiral} with the same expression applied, incorrectly, to the total speed.}
\label{tab:vf}
\centering
\begin{tabular}{lll}
\toprule
$C_L/C_D$ & $e^{-\pi C_D/C_L}$ & simulated $|\vv_f|/|\vv_0|$ \\
\midrule
0.5  & 0.0019 & 0.0435 \\
1.0  & 0.0432 & 0.2110 \\
2.0  & 0.2079 & 0.4622 \\
3.0  & 0.3509 & 0.5938 \\
5.0  & 0.5335 & 0.7263 \\
10.0 & 0.7304 & 0.8506 \\
\bottomrule
\end{tabular}
\end{table}

\begin{figure}[t]
\includegraphics[width=\linewidth]{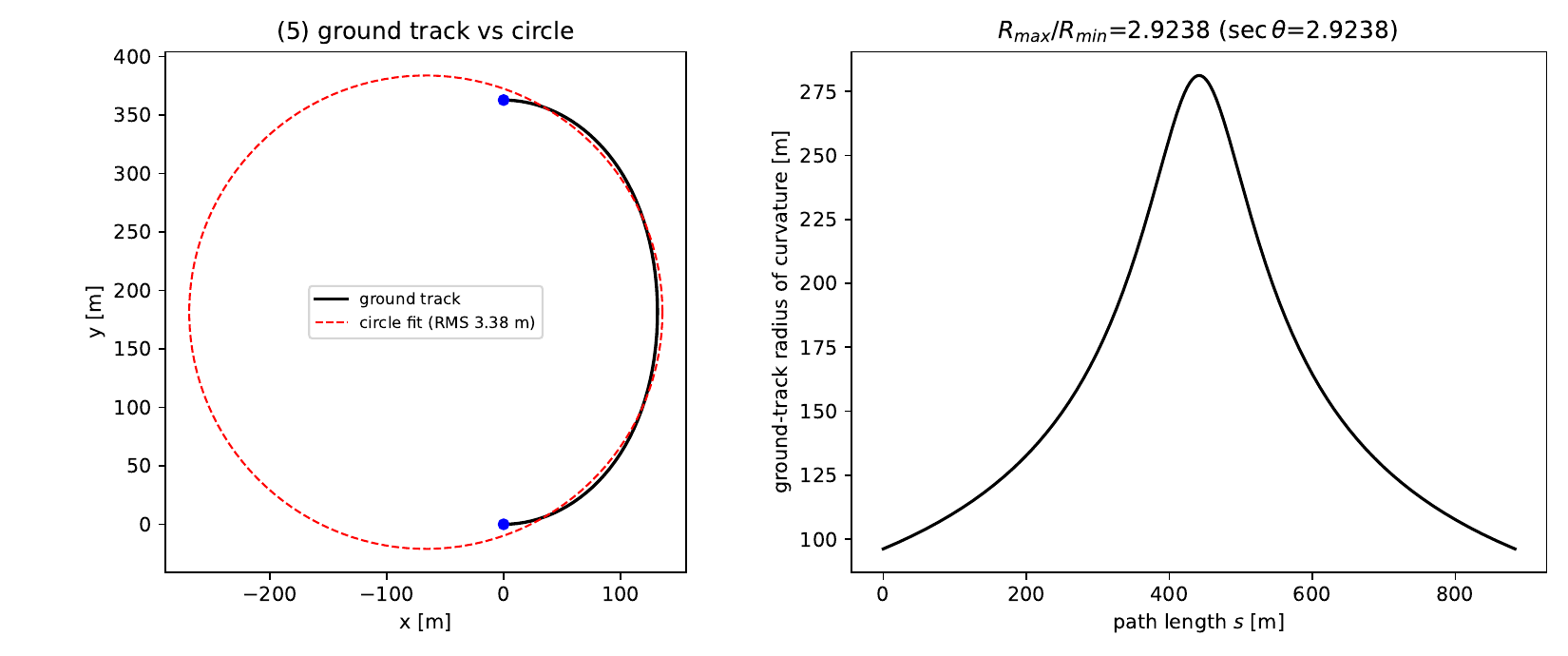}
\caption{Ground track of a drag-free closure solution at $\theta=45^\circ$. Left: the
track against its best-fit circle, with an RMS residual of $1.585$~m on a fitted radius
of $264.06$~m, which is $43$ times the ball radius. Right: the radius of curvature
along the path, following $R_h=\LL\cos\gamma$ with $R_{\max}/R_{\min}=\sec\theta$.}
\label{fig:shape}
\end{figure}

\begin{figure}[t]
\includegraphics[width=\linewidth]{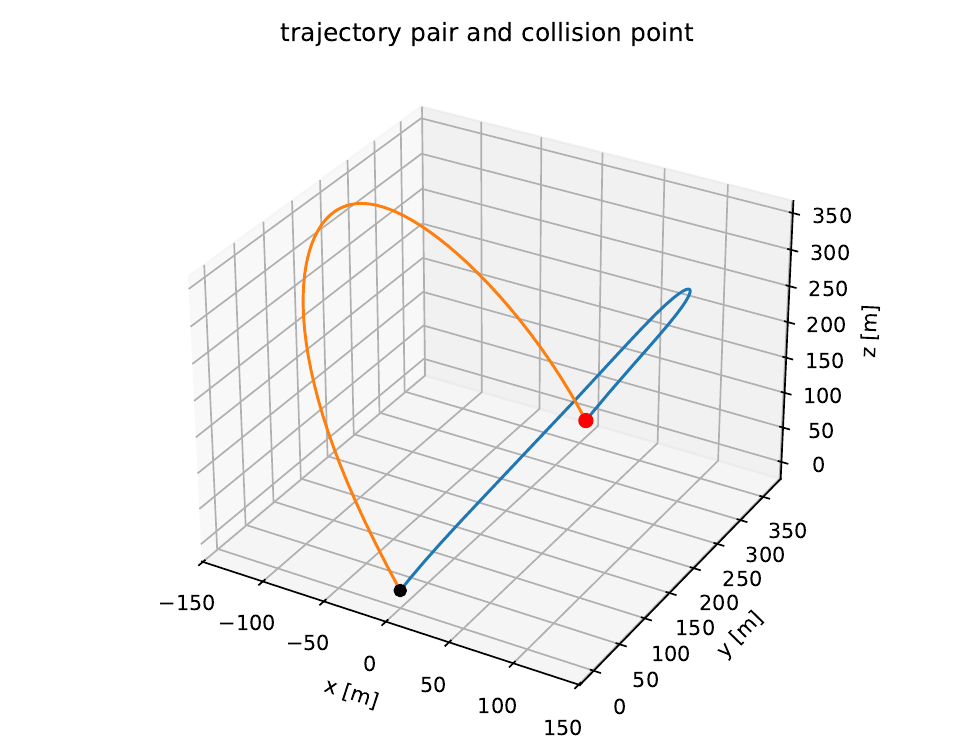}
\caption{The drag-free rendezvous, showing the two counter-spinning trajectories and
their common endpoint at $\theta=45^\circ$ for a baseball. The mirror symmetry is
exact, with arrival times agreeing to $3.6\times10^{-15}$~s.}
\label{fig:pair}
\end{figure}

\section{Discussion}
\label{sec:discussion}

\subsection{Relation to known results}

The individual ingredients used here are old. The complex substitution $w=v_x+iv_y$ is
standard for rotating-frame problems. The constant-$C_L$ turn radius $2m/\rho C_L A$ is
the aircraft-performance result that a wing at fixed lift coefficient turns with a
speed-independent radius. The reparametrisation of projectile motion by a geometric
angle is Bernoulli's, dating to 1719 \cite{Lubarda2022}. What appears not to have been
noticed is that these combine, for the vertical-axis case specifically, into an exactly
solvable system. The turning angle is an affine function of path length, so Bernoulli's
implicit solution becomes explicit, and the turn radius survives the addition of drag
untouched.

It is worth being precise about why the vertical-axis case is special while the general
one is not. For a general spin orientation the Magnus term is not a pure rotation of
$w$, so the horizontal equation does not close and Eq.~\eqref{eq:master} fails. This is
consistent with the perturbative status of the general problem in
Ref.~\cite{Turkyilmazoglu2020}. In exterior ballistics the situation is different
again. For spin-stabilised projectiles the Magnus \emph{force} is usually small enough
to neglect, and the relevant lateral deflection arises instead from the Magnus moment
through the yaw of repose \cite{McCoy2012}, a gyroscopic mechanism unrelated to the one
considered here.

The Magnus--Lorentz analogy familiar from vortex dynamics \cite{Sonin1997} is
instructive precisely where it breaks. There the Magnus force is linear in velocity,
exactly like the Lorentz force, and the resulting circular orbits have a radius
proportional to speed. The aerodynamic Magnus force is quadratic, and the radius is
$\LL$ regardless of speed. That speed-independence is what makes
Eq.~\eqref{eq:pathlength} possible.

\subsection{An open problem}
\label{sec:open}

If one drops the requirement that each ball turn through exactly $180^\circ$ and asks
only that two oppositely thrown balls collide head-on at release height, then the
head-on condition becomes
\begin{equation}
\Delta\psi_A+\Delta\psi_B = 2\pi ,
\label{eq:sumturn}
\end{equation}
so the \emph{sum} of the turns is fixed while the split need not be equal. The equal
split is a drag-free accident. Counting gives four unknowns $v_{0A}$, $\theta_A$,
$v_{0B}$ and $\theta_B$ against four conditions, which suggests isolated solutions with
one ball turning past $180^\circ$. We searched for these over a $70\times70$ grid with
pairing restricted to $\Delta\psi_A<\pi<\Delta\psi_B$, and we found candidates with
genuinely unequal turns of $145^\circ$ and $215^\circ$. Every refinement drained back to
the degenerate corner $\theta\to0$, $v_0\to\infty$ and stalled at a residual miss of
$1.5$--$2.0$~mm, consistent with the $\theta^2$ law of Eq.~\eqref{eq:deficit}. We were
unable to establish either existence or non-existence of a finite-angle asymmetric
solution, and we record it as open.

\subsection{Classroom use}

Several of these results are accessible as exercises at intermediate undergraduate
level, in a first course on classical mechanics that has already introduced quadratic
drag. Deriving Eq.~\eqref{eq:master} from Eq.~\eqref{eq:eom} requires only the
observation that $\omhat\times\vv$ is a rotation, and it makes a clean illustration of
why choosing the right independent variable matters. The statement that a $180^\circ$
turn costs a fixed distance regardless of how hard the ball is thrown is
counter-intuitive enough to be worth setting as a prediction before the derivation. The
$g=0$ circle of Table~\ref{tab:circle} is a good computational exercise, as is
verifying that drag does not appear in Eq.~\eqref{eq:turnrate}. Finally,
Sec.~\ref{sec:tempting} is a useful cautionary example of a plausible pointwise
argument that fails, with the failure diagnosable by plotting one function.

\section{Conclusion}
\label{sec:conclusion}

For a ball spinning about a vertical axis with constant lift and drag coefficients, the
horizontal equation of motion becomes linear with constant coefficients when path
length replaces time as the independent variable, and gravity drops out of it entirely.
The heading then turns at the fixed rate $\kL$ per unit path length, so a given change
of heading always costs a fixed distance. The horizontal speed decays exponentially in
the turn angle at a rate set only by $C_D/C_L$, and the ground-track radius of
curvature is $\LL\cos\gamma$, which reduces to $\LL$ at the apex. Each of these gives a
quantity that can in principle be measured from a single tracked trajectory without
knowing the mass, the radius or the air density separately. The remaining vertical
motion reduces exactly to the one scalar equation \eqref{eq:reduced} in the two
parameters $\lambda$ and $\mu$, which makes the whole flight problem cheap to survey.

Applying this to the symmetric two-ball rendezvous, we find that the meeting is
possible without drag on a one-parameter family given in closed form, and that drag
destroys it. The chord tips below perpendicular by
$(8/3-24/\pi^2)\mu\theta^2$ at leading order, and the sign of that coefficient comes
down to $\pi^2>9$. The asymmetric version of the problem and a proof of the deficit's
positivity over the whole parameter range are both left open.

\section*{Author declarations}

\noindent\textbf{Conflict of interest.} The author has no conflicts of interest to
disclose.

\medskip
\noindent\textbf{Funding.} This work received no external funding.

\medskip
\noindent\textbf{Data availability.} The complete source code, regression test suite
and figure-generating scripts are openly available \cite{ThisCode}. All numbers quoted
in this paper are reproduced by a single driver script, and the analytic claims are
encoded as regression tests with the tolerances stated in Table~\ref{tab:verify}.

\appendix

\section{Leading-order deficit}
\label{app:pert}

We sketch the calculation leading to Eq.~\eqref{eq:deficit}, working to first order in
$\mu$ and second order in $\theta$. For small $Q$, Eq.~\eqref{eq:reduced} linearises to
$Q'\simeq-\lambda(1+2\mu u)$, so
\begin{equation}
Q(u) \simeq \theta - \lambda\bigl(u+\mu u^2\bigr) .
\label{eq:Qpert}
\end{equation}
Imposing closure \eqref{eq:closure}, which at this order reads $\int_0^\pi Q\,du=0$,
gives
\begin{equation}
\lambda = \frac{\theta}{\pi/2+\mu\pi^2/3}
        \simeq \lambda_0\Bigl(1-\frac{2\pi\mu}{3}\Bigr),\qquad
\lambda_0=\frac{2\theta}{\pi} .
\label{eq:lampert}
\end{equation}
With $\cos\gamma\simeq1-Q^2/2$ and $\int_0^\pi\cos u\,du=0$,
\begin{equation}
\mathcal{I} \simeq -\tfrac12\int_0^\pi Q^2\cos u\,du .
\label{eq:Ipert}
\end{equation}
Using $\int_0^\pi u\cos u\,du=-2$, $\int_0^\pi u^2\cos u\,du=-2\pi$ and
$\int_0^\pi u^3\cos u\,du=12-3\pi^2$, and expanding to first order in $\mu$, the terms
independent of $\mu$ cancel identically, as they must by Eq.~\eqref{eq:antisym}. What
is left is
\begin{equation}
\int_0^\pi Q^2\cos u\,du \simeq
 \mu\theta^2\Bigl(\frac{16}{3}-16+\frac{96}{\pi^2}\Bigr) .
\label{eq:Q2int}
\end{equation}
The imaginary part of Eq.~\eqref{eq:chord} is $\int_0^\pi\cos\gamma\sin u\,du\simeq2$ at
leading order, so
\begin{equation}
90^\circ-\beta \simeq \frac{\mathcal{I}}{2}
 = \mu\theta^2\Bigl(\frac{8}{3}-\frac{24}{\pi^2}\Bigr) ,
\label{eq:deficitapp}
\end{equation}
which is Eq.~\eqref{eq:deficit}. The coefficient is positive because $\pi^2>9$.

\bibliography{refs}

\end{document}